\documentclass[fleqn,10pt,a4paper]{SelfArx} 

\usepackage[english]{babel} 

\usepackage{lipsum} 
\usepackage{multirow}
\usepackage{booktabs}
\usepackage{xcolor}
\usepackage{balance}
\usepackage{comment}
\usepackage{caption}
\usepackage{subcaption}
\PassOptionsToPackage{hyphens}{url}\usepackage{hyperref}

\definecolor{color1}{RGB}{0,0,90} 
\definecolor{color2}{RGB}{0,20,20} 

\usepackage{hyperref} 
\hypersetup{hidelinks,colorlinks,breaklinks=true,urlcolor=color2,citecolor=color1,linkcolor=color1,bookmarksopen=false,pdftitle={Title},pdfauthor={Author}}
\usepackage{xspace}

\JournalInfo{\includegraphics[width=3cm]{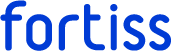}} 
\Archive{fortiss GmbH\\ White Paper, 2020} 

\PaperTitle{Security Engineering for ISO 21434}

\Authors{Yuri Gil Dantas, Vivek Nigam and Harald Ruess\\\small{\textit{fortiss GmbH, Munich, Germany}}\\\small{\textit{lastname@fortiss.org}}} 

\newcommand{\keywordname}{Keywords} 

\newcommand\ie{\emph{i.e.}\xspace}
\newcommand\eg{\emph{e.g.}\xspace}

\newcounter{example}[section]

\newtheorem{SampleEnv}{Example}[section]

\Abstract{
The ISO 21434 is a new standard that has been proposed to address the future challenges of automotive cybersecurity. 
This paper takes a closer look at the ISO 21434 helping engineers to understand the ISO 21434 parts, the key activities to be carried out and the main artefacts that shall be produced. 
As any certification, obtaining the ISO 21434 certification can be daunting at first sight. 
Engineers have to deploy processes that include several security risk assessment methods to produce security arguments and evidence supporting item security claims. 
This paper proposes a security engineering approach that can ease this process by relying on Rigorous Security Assessments and Incremental Assessment Maintenance methods supported by automation. 
The paper demonstrates by example that the proposed approach can greatly increase the efficiency to produce artefacts, as well as enable continuous security assessment.
Finally, the paper points out some key research directions to fully realize the proposed approach.
}
\begin{document}
\sloppy

\flushbottom 

\maketitle 

\tableofcontents 

\thispagestyle{empty} 

\section*{Introduction} 
\addcontentsline{toc}{section}{Introduction} 

New vehicle business models are in the process of being forged based on the 
adoption of Information and Communication Technologies (ICT), such as V2X communication and artificial intelligence.
According to Juniper Research~\cite{juniper}, there are currently tens of millions of connected vehicles, and this number will reach 775 million vehicles by 2023.
Connected vehicles already enable important services, such as on-line streaming and Advanced Driver Assistance Systems (ADAS).
In the near future, they will enable services such as vehicles with very high level of autonomy.

The adoption of ICT, however, has also greatly increased vehicle cybersecurity concerns. 
Until some years ago, attackers would necessarily have to be physically close to carry out attacks against vehicles.
This is no longer the case with connected cars.
As already demonstrated by the now infamous Jeep attack~\cite{attack.jeep}, without appropriate countermeasures, 
attackers can remotely take control of pretty much any vehicle function, such as the vehicle's braking system. 
The year of 2019 has seen an emergence of cyber-attacks~\cite{upstream}: 51 incidents have been reported only in the first quarter of 2019. 
This number is an increase of more than 300\% to 2018. 
Moreover, while in the past attacks were carried out in its majority by white-hats\footnote{Security experts that carry out attacks to find vulnerabilities with the intent of improving security.}, 
now the majority of the attacks reported are carried out by black-hats\footnote{Cyber-criminals that carry out attacks for normally financial gains or notoriety.}
with a rise of 72\% from 2018. 
There is much reason to believe that this trend will continue and even accelerate.
Indeed, as reported by the Hornetsecurity Security Lab~\cite{hornetsecurity}, the automotive industry is the third most targeted by attackers, behind only the energy and logistics sectors.

The cost of handling attacks, \eg, car re-calls, updates, mitigate damage to reputation, will also increase as more connected vehicles enter the roads. Current estimations predict that the cost of attacks could reach USD 23 billion by 2023~\cite{upstream2020}.  Therefore, it is increasingly important that 
vulnerabilities are found and mitigated before production and that new vulnerabilities discovered after production are dealt within a short period of time. 


To address these concerns, the new ISO 21434~\cite{iso21434} has been developed replacing the SAE J3061~\cite{saej3062}. 
It is expected that the ISO 21434 will follow the success of the ISO 26262~\cite{iso26262} for vehicle safety, but now for security.
One goal of this paper is to explain the ISO 21434 providing to cybersecurity engineers a comprehensive overview of the main activities that shall be carried out and the key artefacts that shall be produced to comply with the ISO 21434. 

This paper also goes beyond the ISO 21434.
It describes an incremental approach for security for enabling the continuous security assessment of vehicles and therefore, enabling the continuous certification of vehicles. 
This approach is based on three key elements:
\begin{itemize}
   \item \textbf{Rigorous Security Assessments:} 
   Vulnerabilities are introduced or not correctly mitigated because of the lack of rigor in the security argument used to support a claim. 
   For example, arguments often contain implicit assumptions which lead to unintended implementation errors.
   For the construction of such arguments, this paper suggests rigorous security arguments written by logic-based methods supported by domain-specific languages.
   Logic provides precise semantics specifying, for example, the reasoning principles used to construct safety and security arguments and more importantly the means to check their correctness and completeness.

   \item 
   \textbf{Incremental  Assessment Maintenance:}
   As new vulnerabilities and attacks are constantly being discovered, security updates will be a norm for connected vehicles. 
   The question that follows is how to demonstrate that after an update, a vehicle still complies with the ISO 21434 certification. That is, its security is supported by a valid security argument and appropriate evidence. 
   Re-doing the whole certification process from scratch whenever there is an update is not practical. 
   A better approach is to modify the existing security assessments and certification artefacts incrementally by modifying only the pieces that have been affected by an update.

   \item \textbf{Automation}:
   Increasing the automation of vehicle's security assessments is a necessary requirement for efficient continuous vehicle security assessment and certification.
   Moreover, it also less prone to human-error. 
   Fortunately, there has been in the past year a great increase in the maturity of automated reasoning and verification tools. 
   For example, existing logic programming tools can automatically generate correct safety and security arguments~\cite{dantas20iclp}\cite{dantas20safecomp}, collect~\cite{cruanes13vmcai} and maintain evidence generated from workflows using automated tools~\cite{beyene18fm}.
 \end{itemize} 

\paragraph{Benefits.}
The benefits of the proposed approach with respect to existing approaches are two-fold. 

Firstly, rigorous security assessments with precise semantics enable the automated checking of when an argument is incomplete, \eg, some threats are not appropriately mitigated, or plain wrong. 
Moreover, since the methods are based on logic, engineers can query specifications like the following ones: 
\begin{itemize}
  \item Have all the identified threats been adequately mitigated? If not, which ones remain?
  \item Which architecture options with additional countermeasures can I use to mitigate the threats that have not yet been mitigated? 
  \item What are the impacts of a particular design to other issues such as safety and performance? Will the new countermeasure affect some safety assumption, \eg, increased communication latency?
  \item What is the security argument used to support the security claim of an item?
\end{itemize}

Secondly, the use of incremental methods and automation greatly increases the efficiency of re-certification in terms of time and cost. 
Efficiency is key for enabling continuous certification. 
Otherwise, the overhead costs of re-certification are very high
without adequate automation. 
This has been observed, for example, in the avionics domain, where the change of a single line of code can imply costs of in the order of millions of USD.\footnote{Stated by the DARPA ARCOS project coordinator~\url{https://www.youtube.com/watch?v=bXuBm_awZvg}}

The target audience for this paper is all cybersecurity engineers who are interested in the ISO 21434 itself and in an approach that enables the continuous certification for automotive vehicles. 
It is expected that, by reading this paper, cybersecurity engineers will be able to obtain a clearer picture of the ISO 21434, especially of the activities performed during the risk assessment. 
It is also expected that cybersecurity engineers will both learn that parts of the ISO 21434 may be automated by suitable techniques as well as get a perspective on how to build continuous certification for automotive vehicles using an incremental approach.

Section~\ref{sec:iso-overview} provides an overview of the ISO 21434, and Section~\ref{sec:risk-assess-concept-phase} focuses more on the parts related to risk assessment and concept phase. 
Section~\ref{sec:roadmap} describes an approach for the continuous vehicle assessment and certification.
Section~\ref{sec:approach-by-example} demonstrates this approach through an illustrative example described in Section~\ref{sec:running-example}.
After a discussion of related work in Section~\ref{sec:comparison}, Section~\ref{sec:next-steps} lists the key research directions to fully realize the proposed approach. 
The paper is concluded in Section~\ref{sec:conclusion}.

\begin{figure*}[t]
\begin{center}
  \includegraphics[width=0.9\textwidth]{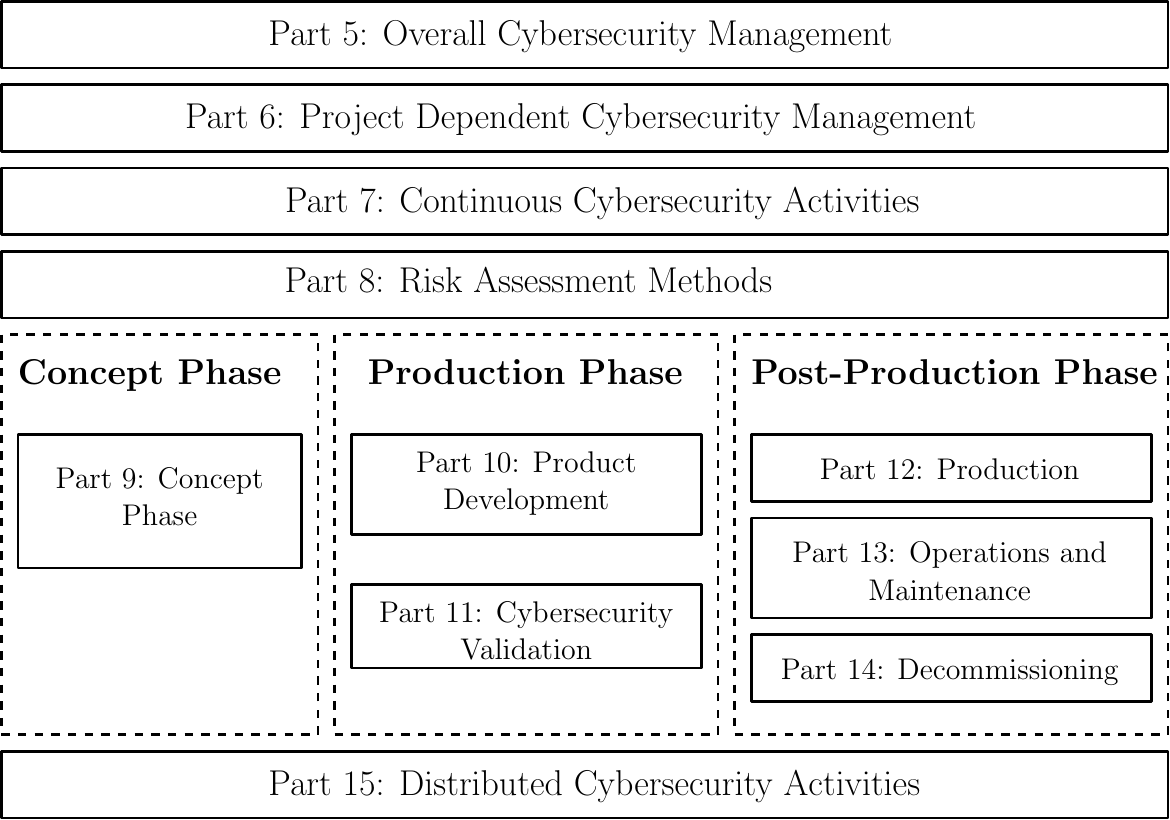}  
\end{center}
  \caption{Technical parts of the ISO 21434.}
  \label{fig:iso-parts}
\end{figure*}

\section{ISO 21434 at a Glance}\label{sec:iso-overview}
As with the ISO 26262, the ISO 21434 is structured into multiple parts (a.k.a. clauses) that tackle different aspects of cybersecurity.
The technical parts of the ISO 21434 are depicted in Figure~\ref{fig:iso-parts}. These parts are preceded by parts defining the scope (Part 1), the normative references (Part 2), the terms and abbreviations (Part 3) and general considerations (Part 4).

Figure~\ref{fig:iso-parts} provides an overview of all the technical parts. 
Section~\ref{sec:risk-assess-concept-phase} goes into more detail regarding Parts 8 (Risk Assessment Methods), and 9 (Concept Phase).  


In a nutshell, the Parts 5 and 6 contain the requirements for organizational cybersecurity. Parts 7 and 8 
describe activities and methods that can be used during the lifecycle of a product to ensure the security of items.
Part 9 details the requirements during concept phase. 
Parts 10 and 11 detail the requirements during production, while Parts 12-14 detail the requirements during the post-production. Finally, Part 15 details the requirements for supplier management.

\paragraph{Part 5.} This part details the requirements 
  for building and maintaining a cybersecurity culture and governance. 
  This is accomplished by setting cybersecurity policies, rules and processes for overall cybersecurity management and for project dependent cybersecurity management. The policies and rules for the overall cybersecurity management shall (1) define the organization-specific rules and processes for cybersecurity, \eg, the executive management's commitment to manage security risks; (2) assign responsibilities for security activities; (3) support the implementation of cybersecurity; (4) institute and maintain cybersecurity culture; (5)
  perform organizational cybersecurity audits; (6)
  manage the sharing of cybersecurity information; (7)
  institute and maintain management systems that support cybersecurity activities; (8) and provide evidence that the tools used do not adversely affect cybersecurity.

\paragraph{Part 6.} This part describes the requirements for the management of cybersecurity activities for the development of a specific project. The key objectives of these requirements are (1) assign the responsibilities for cybersecurity activities;
(2) plan the cybersecurity activities;
(3) create a cybersecurity case that provide the high-level argument for the achieved degree of cybersecurity.

\paragraph{Part 7.} This part details cybersecurity activities that can be performed at any stage of the lifecycle. The main objectives of these requirements are (1) to collect relevant cybersecurity information; (2) to triage collected cybersecurity information; (3) to assess cybersecurity events; (4) to identify and analyze cybersecurity vulnerabilities;
(5) to manage identified cybersecurity vulnerabilities.

\paragraph{Part 8.} This part details methods that organizations can use to determine the risks to stakeholders of cybersecurity threats.
The main objective of the methods are to (1) identify assets; (2) identify threat scenarios; (3) rate the impact; (4) identify or update the attack paths for threat scenarios; (5) assess the ease with which the identified attack paths can be exploited; (6) determine the risk value of a threat scenario; and (7) select the appropriate risk treatment.

\paragraph{Part 9.} This part details the requirements for the concept phase. Its main objectives are 
(1) define the item, the operational environment and its interaction with other items; (2) specify cybersecurity goals and cybersecurity claims; and 
(3) specify cybersecurity requirements and allocate them to the item or to the operational environment.

\paragraph{Part 10.} This part details the requirements for the product development. The main objectives of this part are to (1) specify refined cybersecurity requirements and  architectural design; 
(2) verify that the refined cybersecurity requirements and architectural design comply with cybersecurity requirements from a higher level; 
(3) identify vulnerabilities in the design and manage them accordingly;
(4) provide evidence that the component complies with the cyberscurity specification and does not contain undesired functionality regarding cybersecurity.

\paragraph{Part 11.} This part details the activities for cybersecurity validation that are performed after the integration of components. Its main objectives are 
(1) to validate the cybersecurity goals and claims; (2) to confirm that the item satisfies the cybersecurity goals; (3) to confirm that the residual risk is acceptable.

\paragraph{Part 12.} This part details the requirements during the fabrication, assembly and configuration of an item or component. Its objectives are (1) to apply the cybersecurity requirement for post-development to the item or component; and (2) to prevent the introduction of vulnerabilities during production.

\paragraph{Part 13.} This part describes the requirements for cybersecurity incident response and updates. Its objectives are (1) to handle cybersecurity events that become a cybersecurity incident; and (2) to preserve cybersecurity during and after updates to items or components post-production.

\paragraph{Part 14.} This part details the requirements on decommissioning of an item or component. The main objective is to ensure that the item or component is decommissioned in a secure manner.

\paragraph{Part 15.} This part details the requirements for distributed cybersecurity activities. 
The main objective of these requirements is to define the interactions, dependencies, and responsibilities between customers and suppliers for cybersecurity activities.

\bigskip

Section~\ref{sec:roadmap} describes the proposed approach and how this approach can help to achieve the objectives of Parts 8, 9 and 10 in an automated fashion.
ISO 21434 has several appendixes that illustrate how the ISO 21434 could be used in practice.
Section~\ref{sec:approach-by-example} illustrates the proposed approach using the running example in the ISO 21434, Annex G.
This running example is described in Section~\ref{sec:running-example}.

\begin{figure*}[t]
\begin{center}
  \includegraphics[width=0.9\textwidth]{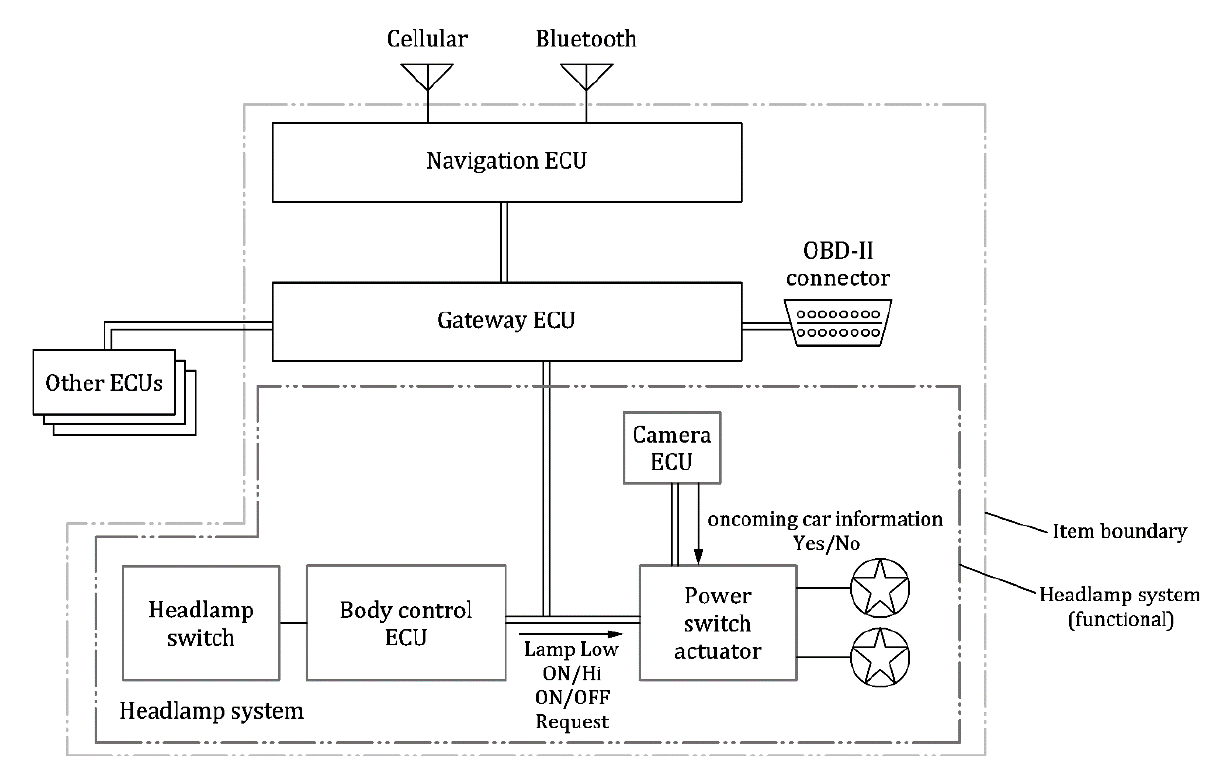}  
\end{center}
  \caption{Preliminary architecture of the headlamp system~\cite{iso21434}}
  \label{fig:headlamp-architecture}
\end{figure*}

\section{Running Example}\label{sec:running-example}
This paper considers a headlamp system as the running example.
The example is taken from ISO-21434, Annex G~\cite{iso21434}.

A headlamp system is responsible for turning on/off the headlamp of a vehicle following the demand of the driver.
It has two specific features, namely high-beam light and low-beam light. 
If the headlamp is in high-beam mode, the headlamp system switches the headlamp automatically to
the low-beam mode when an oncoming vehicle is detected. 
It returns automatically the headlamp to the high-beam mode if the oncoming vehicle is no longer detected~\cite{iso21434}. 
The headlamp system is a critical system as harm, \eg, accidents, may occur if the lamp is unexpectedly turned off over night.

A preliminary architecture of the headlamp system is depicted in Figure~\ref{fig:headlamp-architecture}.
The Body Control ECU sends messages to the Power Switch Actuator through the CAN bus.
These messages are requests to turn on/off the headlamp.
A Camera ECU is connected to the Power Switch Actuator so that oncoming vehicles can be detected.
This is needed to automatically switch the headlamp from high-beam mode to low-beam mode and vice versa.
The CAN bus is connected with a Gateway ECU.
This Gateway ECU controls access to the CAN bus from other ECUs located outside the headlamp system such as the Navigation ECU.
The Navigation ECU has two interfaces, namely Cellular and Bluetooth, that may be used to send requests to turn on/off the lamp.
The Gateway ECU is also connected to an OBD-II connector through another CAN bus.
OBD-II is an on-board computer that monitors data about the vehicle such as speed.
Both the Bluetooth and Cellular interfaces and the ODB-II connector are located outside the item boundary\footnote{The description of the item boundary can
include interfaces with other item and components internal and/or external to the vehicle~\cite{iso21434}}.
They might be entry points for attacking the headlamp system, as discussed in the following sections.


\section{Risk Assessment}\label{sec:risk-assess-concept-phase}
This section enters into the details of the technical parts of the ISO 21434, namely the Parts 8 and 9. 
Part 9 (\ie, item definition) may use the risk assessment methodology described in Part 8. 
This methodology can be partially carried out with automated support, thus enabling the automated production of several artefacts required for the security assessment of an item. %

\begin{figure*}
\begin{center}
\begin{center}
  \includegraphics[width=0.95\textwidth
 ]{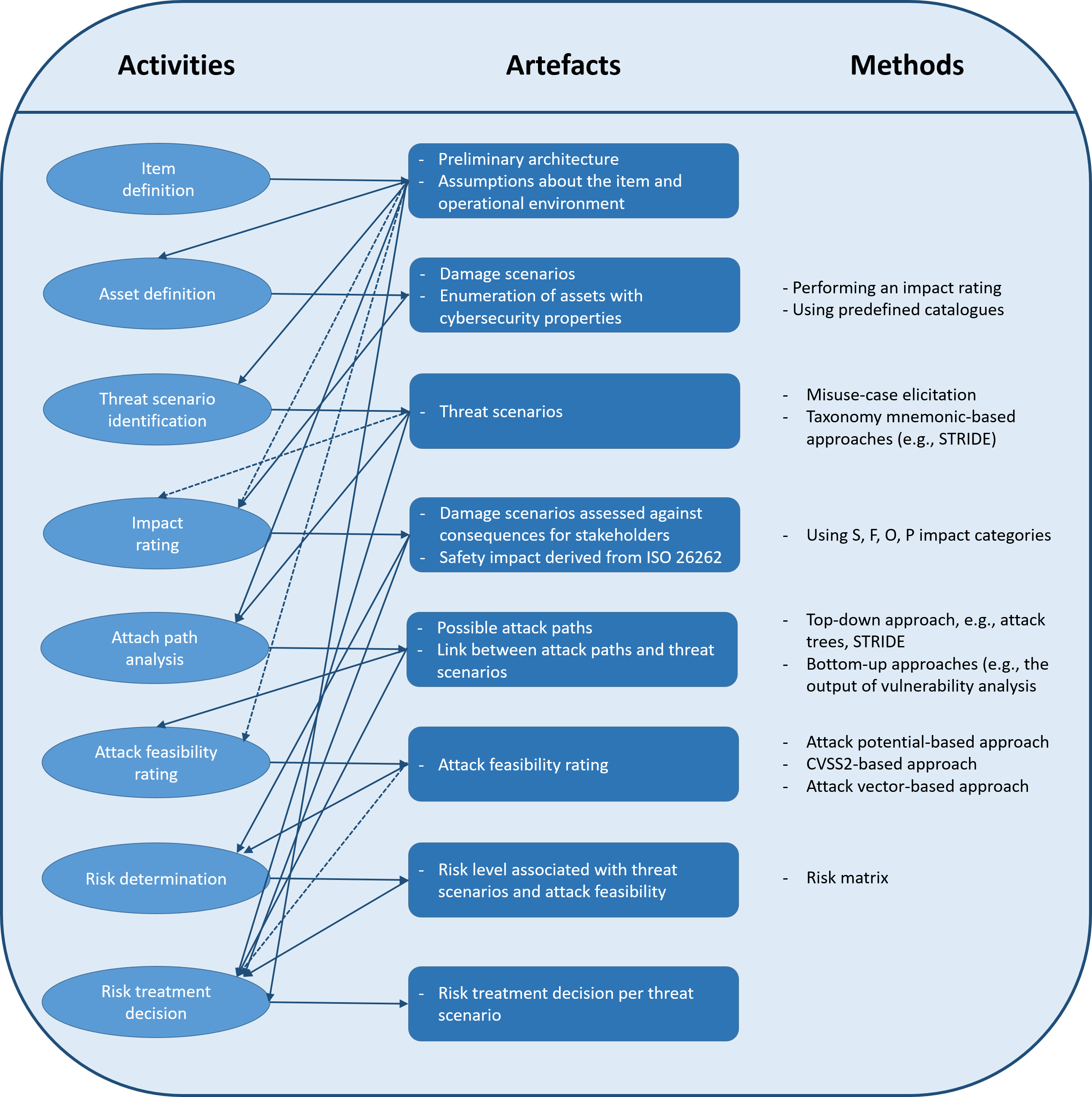}  
\end{center}
\end{center}
  \caption{Activities, Artefacts and Methods mentioned in the Part 8. Item definition is mentioned in the Part 9.
  The full arrow from an activity to an artefact specifies that this artefact is produced at the end of the activity. A full arrow from an artefact to an activity specifies that this artefact is used as input to the corresponding activity. Finally, the dashed arrows from an artefact to an activity specifies that the artefact can used as input, but not required, to the corresponding activity.}
  \label{fig:part8}
\end{figure*}

Figure~\ref{fig:part8} depicts the activities described in Parts 8 and 9 (item definition) of the ISO 21434 for risk assessment.
It includes the artefacts that they are supposed to produce and suggestions of methods that can be used to produce them. 

The following text describes these activities and illustrate them with the running example from Section~\ref{sec:running-example}.

\paragraph{Item Definition.} This is an early-on activity that defines the item\footnote{An item is a system or combination of systems to implement a function.} whose risk assessment is going to be evaluated. 
During this activity, a preliminary architecture of the item is described and the assumptions about its operational environment. 

\begin{SampleEnv}\label{ex:item-identification}
Consider the running example from Section~\ref{sec:running-example}. The artefacts produced by this activity are the preliminary architecture of the headlamp system, the description of its architectural elements (a.k.a. functions), and the item boundary. These artefacts are described in Section~\ref{sec:running-example}.
\end{SampleEnv}


\paragraph{Asset Identification.} From the artefacts produced by the item definition activity, the item's damage scenarios\footnote{A damage scenario is an adverse consequence or undesirable result due to the compromise of a cybersecurity property of an asset} are identified.
Moreover, one should also enumerate the identified assets with cybersecurity properties that would lead to a damage scenario.
Three methods are suggested by the ISO 21434 for carrying out this activity:

\begin{itemize}
  \item Impact rating: This method rates the impact of cyberattacks to assets.
  \item Deriving Assets from Threat Scenarios:\footnote{A threat scenario is a statement of potential negative actions that lead to a damage scenario.}
  Threat scenarios (produced in the threat scenario activity) can help identify key assets.
  \item Pre-defined Catalogues: Existing catalogues provide a good source for identifying assets.
\end{itemize}

\begin{SampleEnv}\label{ex:asset-identification}
An asset from the headlamp is the CAN bus that transmits messages such as requests to turn on/off the lamp.
Cybersecurity properties relevant for the CAN bus include integrity and availability. 
Integrity-wise, the CAN bus must ensure accuracy and completeness of the message transmitted. 
Availability-wise, the CAN bus must be available at all times whenever, \eg, the Body Control ECU requests to turn on/off the lamp.
Damage scenarios for the CAN bus include unexpected behavior like:
\begin{enumerate}
\item \textbf{Unexpected loss of head light:} the headlamp turns off during night driving resulting from the loss of integrity of CAN signal
\item \textbf{Unable to switch on head light:} the Power Switch Actuator does not receive a request from the Body Control ECU to turn on the lamp resulting from the loss of availability of the CAN bus.
\end{enumerate}
\end{SampleEnv}

\paragraph{Threat Scenario Identification.} The threat scenarios for each damage scenario shall be identified. Notice that a damage scenario may be associated with multiple threat scenarios. A threat scenario may include the targeted asset, the compromised cybersecurity property and the actions that need to be carried out by an attacker to accomplish a damage scenario.

While there is a number of methodologies to carry out this activity~\cite{shostack14book}, the ISO 21434 proposes the use of the two following methods or their combination: 

\begin{itemize}
    \item misuse-case elicitation:
    Threats scenario can often be identified by using items in an possible, but unintended way.
    \item taxonomy mnemonic-based approaches, such as STRIDE: Using threat categories enables a systematic identification of threat scenarios. (See~\cite{shostack14book} for more about this type of technique.)
\end{itemize}  

For this activity, there might be the need for an additional artefact specifying what type of attacker is being considered.
This artefact is important as attacker intentions, (\eg, is he a script-kid, cyber-criminal or an agent supported by a government?, and capabilities, \eg, which types of channels is he able to hack?) lead to different types of threat scenarios. 


\begin{SampleEnv}\label{ex:threat-scenario-identification}
Consider the damage scenarios from Example~\ref{ex:asset-identification}.
Threat scenarios for Damage Scenarios 1 and 2 can be, respectively:
\begin{enumerate}
  \item An attacker spoofing a CAN signal leads to loss of integrity of the CAN message of `Lamp request'.
  \item An attacker flooding the CAN bus buffer leads to loss of availability of the CAN bus.
\end{enumerate}

\begin{table*}[ht]
\centering
\begin{tabular}[t]{lcccc}
\hline
Damage scenario &Safety&Financial&Operational&Privacy\\
\hline
Unexpected loss of head light&Major&Negligible&Major&Negligible\\
Unable to switch on head light&Moderate&Negligible&Major&Negligible\\
\hline
\end{tabular}
\caption{Impact rating for the damage scenarios from Example~\ref{ex:asset-identification}.}
\label{table:impact-rating}
\end{table*}%

For the Threat Scenario 2, e.g., an attacker could flood the CAN bus with high priority messages so that the lamp switches off requests may be processed with significant delay.




\end{SampleEnv}

This artefact describing the attacker would enable the use of formal 
approaches to enumerate in an automated fashion threat scenario identification by using precise mathematical models~\cite{dantas20safecomp,dantas20vnc,nigam19etfa,urquiza19csf}.
Section~\ref{sec:approach-by-example} illustrates how automation is a cornerstone for incremental security engineering.

\paragraph{Impact Rating.}
This activity takes as input the damage scenarios derived from the activity Asset Identification. 
Each damage scenario is assessed according to four impact categories for the stakeholders: safety (S), financial (F), operational (O) and privacy (P).
Other categories may be considered. 
Moreover, one of the following impact rating is associated with the damage scenario: severe, major, moderate, or negligible.
Finally, safety related impacts shall be derived from the ISO 26262~\cite{iso26262}. 
The resulting artefact are the damage scenarios assessments.


\begin{SampleEnv}\label{ex:impact-rating}
The impact rating for the damage scenarios from Example~\ref{ex:asset-identification} is described in Table~\ref{table:impact-rating}.

Regarding the first damage scenario, an accident might happen if the lamp switches off overnight driving.
Hence, the safety impact for this scenario is major.
The operational impact is also major, because if the lamp is not working properly the driver needs to take the vehicle to a mechanic for repair.
The damage impact for both financial and privacy categories are negligible for this scenario.

For the second damage scenario, this paper considers that the driver is not able to switch on the lamp over morning driving.
The safety impact is then moderate, as the impact would not be as severe as driving at night.
The operational impact is major, following the same reason explained for the first scenario.
The damage impact on financial and privacy are both negligible for this scenario.
\end{SampleEnv}

While no concrete method is proposed by the ISO 21434 for assessing damage scenarios, this paper points out that there are methods in the literature that can help, in particular, assess damage scenarios related to safety. 
For example, ~\cite{kondeva19wosocer} proposes methods for extracting in an automated fashion security relevant information from safety analysis, \eg, Fault Tree Analysis (FTA), Failure Modes and Effects Analysis (FMEA), and safety cases.

\paragraph{Attack Path Analysis.}
From the preliminary architecture, assumptions about the item/environment, and the damage scenario, 
one produces the possible attack paths and their association to threat scenarios. 
To do so, the ISO 21434 proposes the use, possibly in combination, of both top-down, \eg, Attack Trees, and bottom-up, \eg, vulnerability analysis. 
The attack paths shall include the following information: vulnerabilities or weaknesses that could be exploited, and how these could be exploited in an attack realizing a threat scenario.


\begin{SampleEnv}\label{ex:attack-path-analysis}
Both Bluetooth and Cellular interfaces may be accessible from the outside, possibly by an attacker.
Hence, this paper considers such interfaces as the entry point vulnerabilities for the headlamp system.

Consider the threat scenarios from Example~\ref{ex:threat-scenario-identification}.
A threat path for Threat Scenario 1 can be:
\begin{enumerate}[label=(\alph*)]
  \item an attacker compromises the Navigation ECU from the Cellular interface to send malicious signals to turn off the lamp during night driving,
  \item compromised Navigation ECU transmits the received malicious signals, and
  \item Gateway ECU forwards the malicious signals to Power Switch Actuator through the CAN bus.
\end{enumerate}
Note that this very same attack could be also carried out from the Bluetooth interface.
Similarly, a threat path for Threat Scenario 2 can be:

\begin{enumerate}[label=(\alph*)]
  \item an attacker compromises the Navigation ECU from the Bluetooth interface to flood the CAN bus with high priority messages,
  \item compromised Navigation ECU transmits the received malicious messages, and
  \item Gateway ECU forwards the malicious messages to the CAN bus.
\end{enumerate}
\end{SampleEnv}

\begin{table*}[ht]
\centering
\begin{tabular}[t]{ccccccc}
\hline
attack steps&expertise&time (\#days)&equipment&knowledge&opportunity&level\\
\hline
(a)&expert&14&standard&public information&unlimited&high\\
(b)&expert&7&standard&restricted information&unlimited&high\\
(c)&proficient&1&standard&restricted information& unlimited&high\\
\hline
\end{tabular}
\caption{Attack feasibility for the attack path for Threat Scenario 1 (Example~\ref{ex:attack-path-analysis}).}
\label{table:attack-feasibility}
\end{table*}%

\paragraph{Attack Feasibility Rating.}
Each attack path is rated according to the following categories: high, medium, low or very low. 
The following three methods are suggested for rating attack paths:
\begin{itemize}
  \item Attack Potential: The rating of an attack path is obtained by evaluating the attack level taking into account core factors required, including elapsed time, specialist expertise and item knowledge required, window of opportunity, and equipment.
  \item CVSS2: The attack path rating is obtained from common vulnerability scoring systems, \eg, available at \url{https://www.first.org/cvss/}.
  \item Attack vector: The rating is determined by analysing the predominant attack vector of the attack path. 
\end{itemize}

\begin{SampleEnv}\label{ex:attack-feasibility-rating}
This example uses the attack potential approach.
Table~\ref{table:attack-feasibility} shows the attack feasibility rating for the attack path for Threat Scenario 1 from Example~\ref{ex:attack-path-analysis}.
Note that the chosen parameters in Table~\ref{table:attack-feasibility} are samples only and may not reflect reality.

Expertise is the skill level required to exploit an attack path.
This paper considers that an expert is required to compromise the Navigation ECU though the Cellular interface (attack step (a)). 
It also considers that the total time for an expert to exploit attack step (a) is 14 days.
Equipment refers to hardware or software required to exploit an attack path.
The Cellular interface may be readily available and easily accessible by the attacker though, e.g., wireless network.
Therefore, the equipment required for exploiting attack step (a) is standard.
The knowledge required by the attack is public, as information regarding the Cellular interface may be available on the Internet.
Opportunity is the amount of access required by the attack to exploit an attack path. 
The opportunity is unlimited for step (a), because the Cellular interface may be available without any time limitation to the attacker.

The attack potential approach considers the most expensive argument from each attack step to compute the overall feasibility for the attacker's path.  
For example, if you need an expert for one attack step and a proficient for another attacker step the overall value w.r.t. expertise is expert.
The value for each attack's step is high (see Table~\ref{table:attack-feasibility}).
Hence, the overall attack feasibility for the attack path for Threat Scenario 1 is high.
This paper considers that the overall attack feasibility for the attack path for Threat Scenario 2 is medium.

\end{SampleEnv}




\paragraph{Risk Determination.} The risk of threat scenarios is determined from the impact associated to its corresponding damage scenario and the likelihood of its associated attack paths. 
The risk are integers from 1 (lowest risk) to 5 (highest risk).
A risk matrix with the likelihood of attack paths and impact of the damage scenario is a suggested method to carry out this activity.


\begin{SampleEnv}\label{ex:risk-determination}
One can determine the risk value for both threat scenarios from Example~\ref{ex:threat-scenario-identification} based on the risk matrix.
To this end, consider both the impact rating from Example~\ref{ex:impact-rating} and the attack feasibility rating from Example~\ref{ex:attack-feasibility-rating}.

This paper considers the safety impact rating only from Example~\ref{ex:impact-rating}.
The risk value for Threat Scenario 1 is 4, as the impact on safety is major and the likelihood of its attack path is high.
The risk value for Threat Scenario 2 is 2, as the impact on safety is moderate and the likelihood of its attack path is medium.

\end{SampleEnv}



\paragraph{Risk Treatment Decision.}
From all the artefacts produced by the other activities, excepting the attack feasibility rating which is optional, a risk treatment decision is produced for each threat scenario.
This artefact shall contain treatment options, such as avoiding risk by removing risk sources, reducing risk, by, \eg, inserting countermeasures, sharing or transferring the risk to, \eg, suppliers or through insurance, or accepting or retaining the risk.

\section{Incremental Approach}\label{sec:roadmap}
While the ISO 21434 is a step forward for the cybersecurity of connected vehicles, it does not enter 
into the details of how to \emph{efficiently operationalize cybersecurity}.
Given the complexity of connected vehicles, carrying out all activities and producing all artefacts while still satisfying the production timelines is not feasible without automation. Moreover, cybesecurity is a continuous task as new vulnerabilities and 
attacks are discovered (after vehicle production) requiring new countermeasures and analysis.

This paper proposes an incremental approach for ISO 21434 that enables continuous and efficient cybersecurity based on three main pillars:
\begin{itemize}
  \item \textbf{Rigorous Security Assessments:}
  This paper proposes security rigorous assessment specification based on precise logical models with precise semantics.
  In contrast with existing assessment languages, such as Goal Structure Notation~\cite{gsn11standard}, that include textual elements, rigorous security assessments can be automatically checked for missing assumptions or flaws in the argumentation.

  Following the previous work by~\cite{dantas20safecomp,dantas20iclp,kondeva19wosocer}, this paper proposes domain specific languages enabling the specification of system architectures, \ie, components and channels, including physical components, \eg, CAN bus or ECUs, 
  and safety and security elements, \eg, hazards and threats, as well as safety and security architectural patterns, \eg, safety and security monitors, watchdogs, and firewalls.

  From this language, safety and security reasoning principles are specified as logic programming rules. They specify, for example, under which conditions 
  a particular security pattern used in a given architecture can guarantee the security against some given threats.

  \item \textbf{Incremental Assessment Maintenance:} Whenever there is a change in the system, \eg, a new function is added or removed, or a change in the environment, \eg, a new threat is identified, 
  the security assessment (argument and evidence) shall be re-assessed. 

  A better approach than re-assessing the whole security assessment is to re-assess the parts of the security assessment that are affected by these incremental changes.

  Automated incremental techniques have been developed 
  for logic programming~\cite{gupta93sigmod,nigam12comlan}. 
  These work specifies algorithms for maintaining incrementally (distributed) logic specifications used in databases and for network routing. 
  It seems possible to use and adapt these techniques for the maintenance of rigorous security assessments.

  \item \textbf{Automation:} For increased process efficiency and reduced errors, increased automated support is key. 
  
  Fortunately, the past years have witnessed the emergence of several automated methods for 
  the reasoning about system architectures by, \eg, design exploration, and the generation of evidence by using (formal) tools, such as static analyzers and model-checkers.
\end{itemize}

\begin{figure*}[t]
  \begin{center}
    \includegraphics[width=0.90\textwidth]{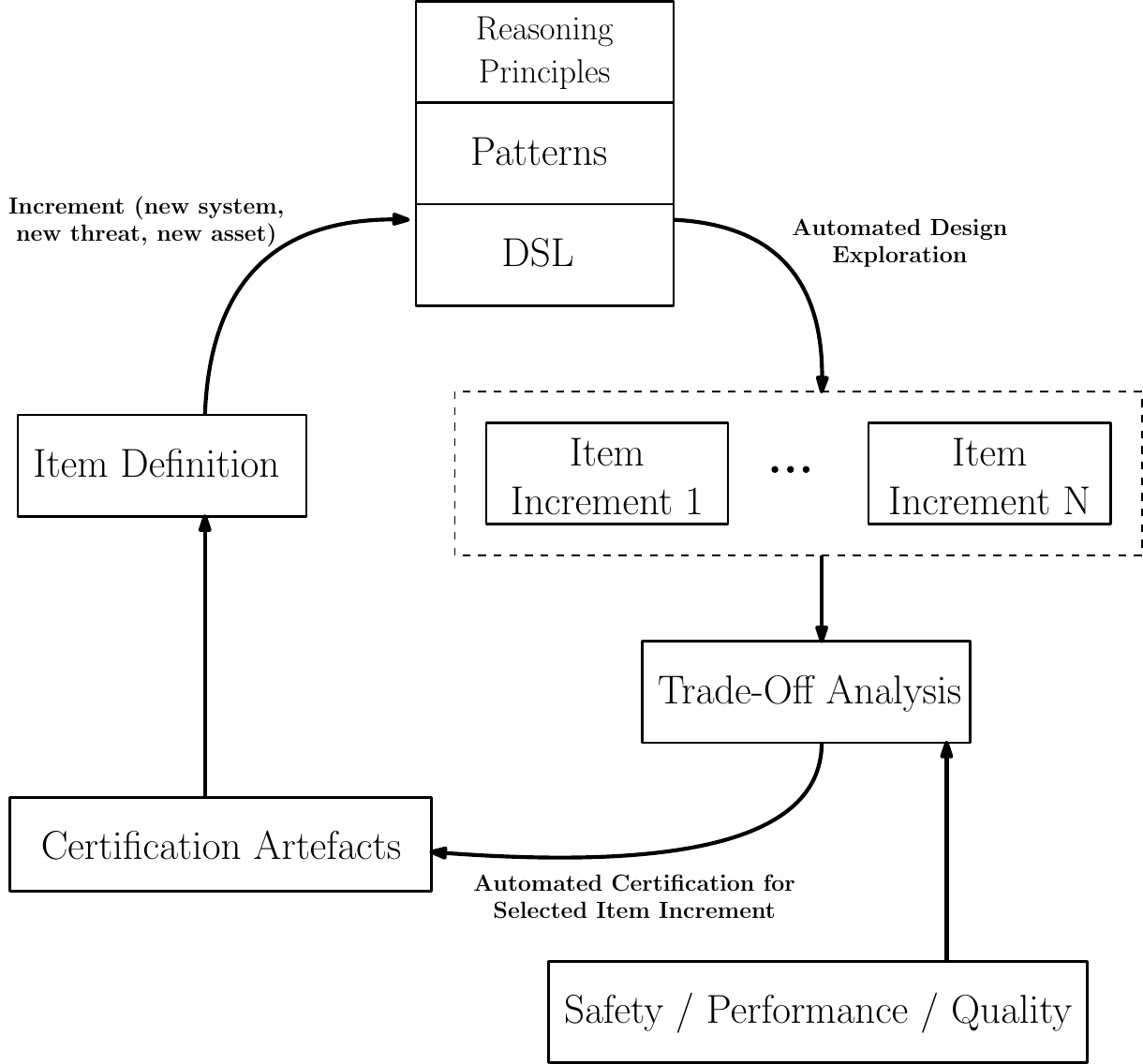}
  \end{center}
  \caption{Incremental Security Approach for ISO 21434.}
  \label{fig:incremental}
\end{figure*}

Figure~\ref{fig:incremental} depicts how the three pillars above are used for the incremental security engineering of ISO 21434.
This paper refers to these three pillars working together as the \emph{machinery}, especially when referring to the automation of tasks. 

Assume a given item definition $I$ for which a rigorous security assessment has been carried out. Moreover, assume that 
there is a new increment, $\Delta$, to the item, \eg, new feature, or to the assumptions about the environment, \eg, new threat, that may invalidate the existing security assessment for $I$.

The increment $\Delta$ is fed to the logic specification based on a Domain-Specific Language (DSL) on which architectural patterns and reasoning rules are specified.
By using design exploration techniques, the machinery automatically generates a collection of item increments all of which are associated with a security assessment. 

A trade-off analysis among the generated item increments taking into account other aspect, such as safety, performance, costs or quality.
This trade-off is supported by automated techniques enabling the selection of a best suited item increment. 

Finally, the certification artefacts, \eg, security assessments, are generated for the new item.

\paragraph{Benefits.}
The proposed approach has two key benefits to existing approaches based on model-based engineering, \eg, GSN models, or textual models, \eg, Excel sheets. 

The first benefit is derived from the use of logic-based specifications for the generation of security assessment.
As a result, one can rely on concepts like soundness and completeness using precise semantics, instead of the textual semantics which may be ambiguous in GSN models for example.
This enables the automatic checking of whether the argument used in security assessments is correct or whether there are flaws in the reasoning or missing assumptions. 
Moreover, using logic-based tools, it is even possible to automatically enumerate which evidence is still required for completing a security argument.

The second benefit is provided by the use of incremental methods and automation. 
This enables a less human error-prone and efficient certification process. 
Without these two elements, the overhead costs for certification is very high as already observed in the avionics domain where the change of a single line of code can imply costs of in the order of millions of USD.

A key impact of the two benefits above is that the proposed incremental approach enables continuous certification of connected vehicles. 
Whenever there is a new incremental change to the item or to the environment assumptions, 
new certification artefacts should be produced for certification.
The use of rigorous security assessments and the incremental approach enable the production of these artefacts in a highly automated fashion by relying on automated reasoning tools, such as logic programming tools~\cite{dlv,cruanes13vmcai}. 
Moreover, certification authorities have to check whether the artefacts are compliant to the ISO 21434. 
Since the methods used in the approach are based on rigorous security assessments, certification authorities can use existing tools to automatically check whether the security arguments are correct.

\section{Approach by Example}\label{sec:approach-by-example}

The goal of this section is to describe (1) how the proposed approach can be used to automate parts of the ISO 21434 risk assessment and (2) how this approach deals with incremental changes in the system architecture, thus supporting continuous security assessment and certification.
To this end, the machinery is applied to the headlamp system explained in Section~\ref{sec:running-example}.

\subsection{Automating ISO 21434}\label{ssec:automating-iso21434}
The machinery takes as input the preliminary system architecture as well as the damage scenarios associated to both assets from the system architecture and cyber-security properties.
The machinery also expects as input the impact rating for each given damage scenario.
These artefacts cover three activities from the risk assessment, namely item definition, asset identification, and impact rating.\footnote{Since these three activities are specific to the product being developed, 
they cannot be automated in general.}

\begin{figure*}[t]
  \begin{center}
    \includegraphics[width=0.90\textwidth]{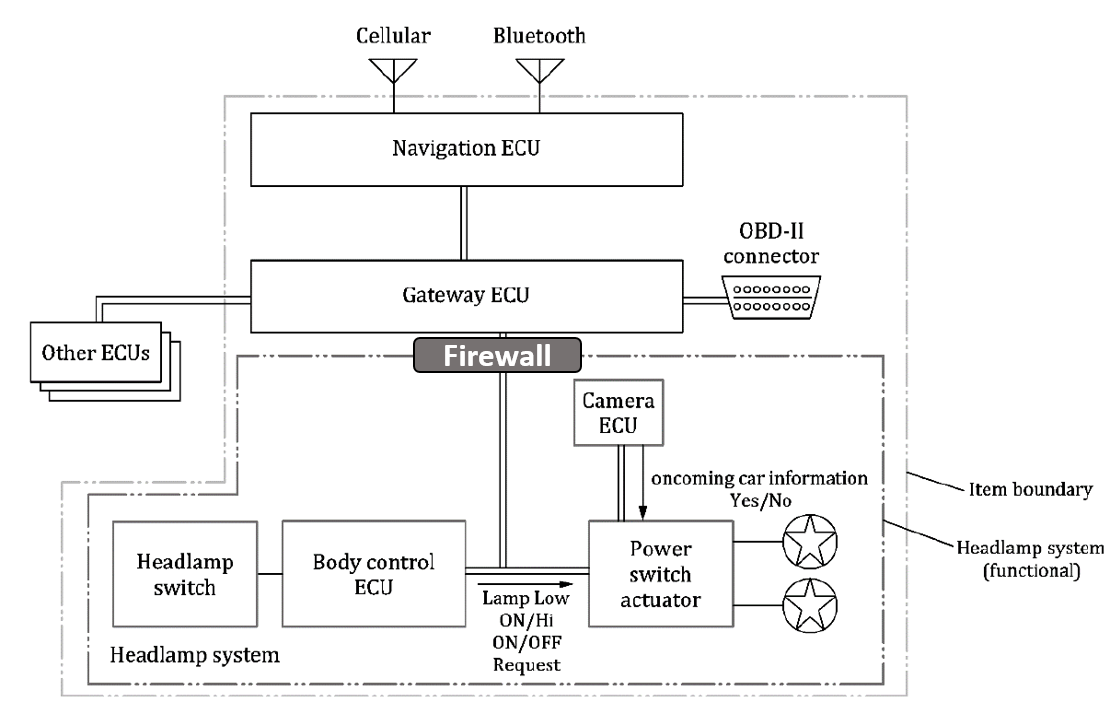}
  \end{center}
  \caption{Headlamp system with a Firewall}
  \label{fig:firewall-security-pattern}
\end{figure*}

Consider the Example~\ref{ex:asset-identification} from Section~\ref{sec:risk-assess-concept-phase}. 
The asset is the CAN bus transmitting messages from the Gateway ECU to the headlamp system.
This section considers the Damage Scenario 1 only, namely unexpected loss of head light.

\begin{SampleEnv}\label{ex:dlv-asset-damage-scenarios}
In the DSL, one can specify both the identified asset and damage scenarios as follows:

\begin{verbatim}
asset(can2). 
dmgScenario("hl turns off 
over night driving",can2,int,
[maj,neg,maj,neg]).
\end{verbatim}
The argument can2 denotes a Controller Area Network (CAN) used to communicate between ECUs. 
The fact asset(can2) denotes that can2 is an asset. 
The fact dmgScenario denotes a damage scenario (headlamp turns off over night driving) associated with can2 and cyber-security property int (short for integrity). 
Following the impact rating given in Table~\ref{table:impact-rating}, the impact rating for this scenario is specified as [maj,neg,maj,neg] (maj and neg are short for major and negligible, respectively). 

Note that this example omits how the system architecture is specified in the DSL.
The interested reader can find detailed description on how to specify elements for the system architecture such as channels and components in~\cite{dantas20iclp}. 
\end{SampleEnv}

The next activities are the threat scenario identification and attack path analysis.
The machinery can automate these activities. 
The reasoning principles specify a potential threat for each asset associated to a damage scenario. 
A potential threat becomes an actual threat if the asset may be reached by an attacker.
The machinery expects as input the public elements of the system architecture, \ie, the elements that may be accessible by external users.
Bluetooth and Cellular interfaces and OBD-II connector are considered to be public elements.
Reachability rules are specified to automatically check whether from a public element (\eg, Bluetooth) an attacker may access an asset (\eg, CAN bus) through a path P. 
The machinery identified three possible paths to access the CAN bus (located between Gateway ECU and headlamp system) from the public elements.

Next in the risk assessment are the attack feasibility rating and risk determination.
As shown in Table~\ref{table:attack-feasibility}, a level (\eg, high) is assigned to each step of an attacker's path.
The machinery takes as input the assigned level to each attack step of a given attack path.
It then automatically calculates the overall level of an attack path following the attack potential approach.
Given the overall level of an attack path, reasoning principles rules are specified to automatically calculate the risk determination for each threat following the risk matrix method.

\begin{SampleEnv}\label{ex:attFS-riskDT}
The machinery outputs the following w.r.t. the attack feasibility rating and risk determination.

\begin{verbatim}
{attFS([can2,[can2,gw,can1,bt],
int,maj],high)}
{riskDT([can2,[can2,gw,can1,bt],
int,maj],4)}
\end{verbatim}

For both facts, attFS and riskDT, the first argument is the threat ID.
The threat ID is composed of the identified asset (can2), the attacker's path (direction from right to left), the cyber-security property (integrity), and the impact rating (major).
This example considers the safety impact rating only.
The last argument of attFS and riskDT denotes, respectively, the overall feasibility rating (high), and the risk value for the identified threat (4).
\end{SampleEnv}




The last activity in the risk assessment is risk treatment decision.
The machinery enables the automated recommendation of security patterns to mitigate identified threats.
To this end, reasoning principles are specified for (1) \emph{Mitigation}: which threats can be mitigated by a given deployed security pattern and which threats cannot be mitigated. 
(2) \emph{Security Pattern Recommendation:} which security patterns, \eg, Firewall or Security Monitor, can be used and where exactly they should be deployed to mitigate threats that have not yet been mitigated.

\begin{figure*}[t]
     \centering
     \begin{subfigure}[b]{0.45\textwidth}
         \centering
         \includegraphics[width=\textwidth]{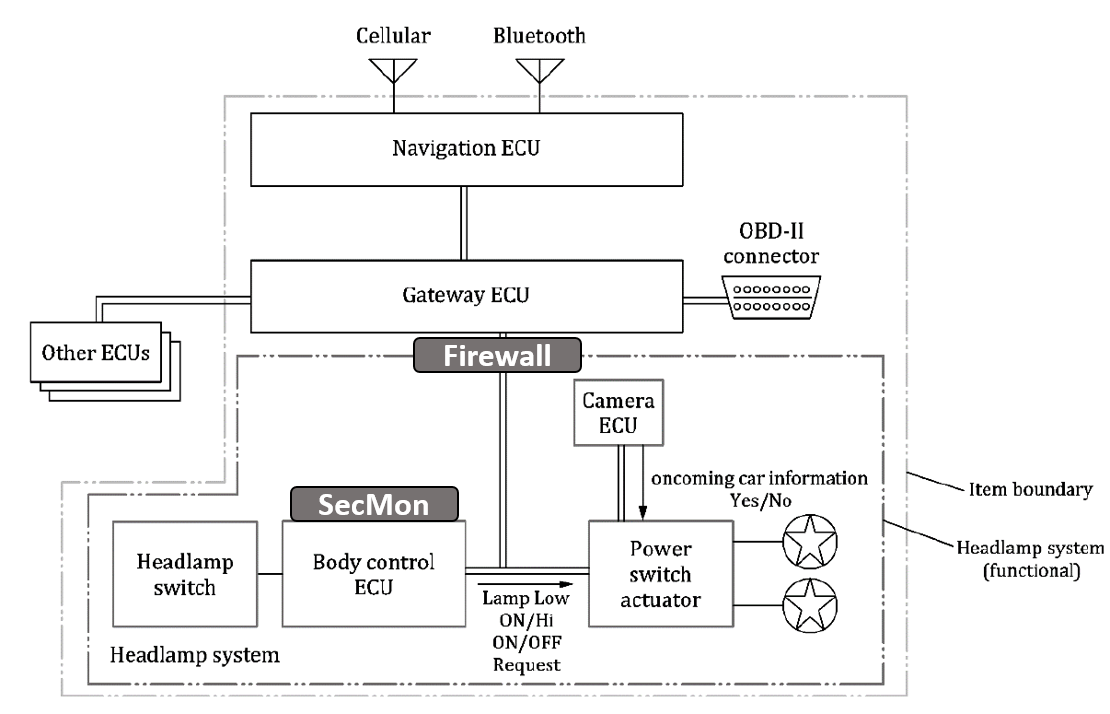}
         \caption{Security monitor associated to the Body Control ECU}
         \label{fig:firewall-secmon-security-patterns-solution1}
     \end{subfigure}
     \hfill
     \begin{subfigure}[b]{0.45\textwidth}
         \centering
         \includegraphics[width=\textwidth]{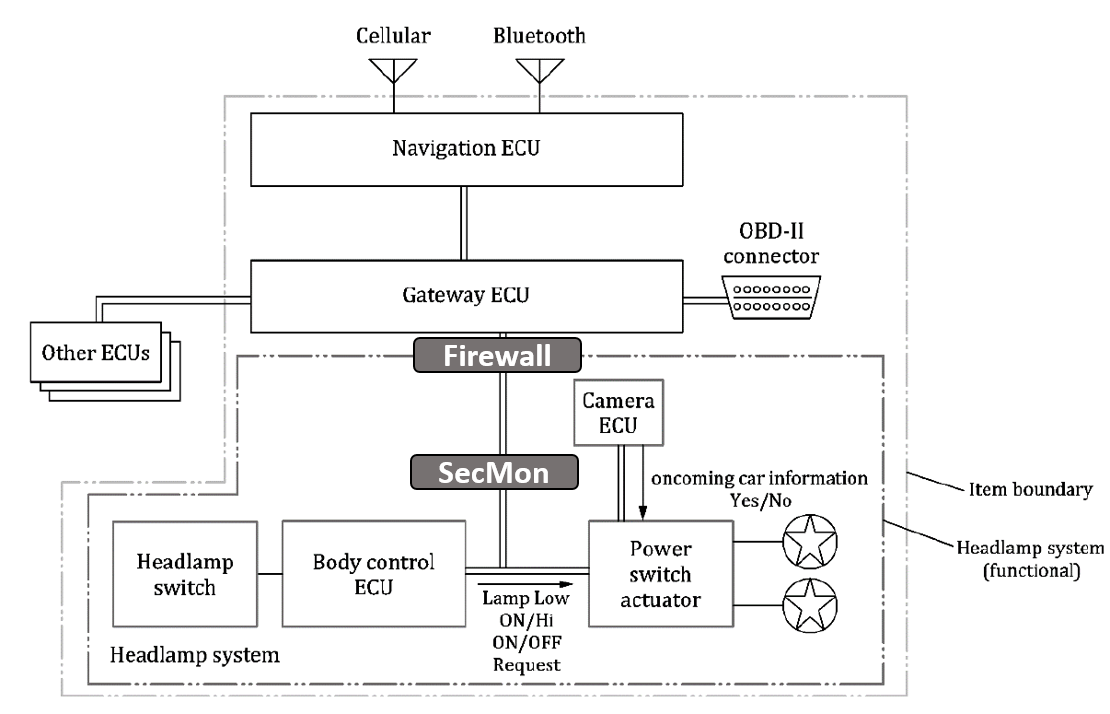}
         \caption{Security monitor associated to the CAN bus}
         \label{fig:firewall-secmon-security-patterns-solution2}
     \end{subfigure}
        \caption{Headlamp system with a Firewall and a Security Monitor}
        \label{fig:firewall-secmon-security-patterns}
\end{figure*}

The machinery recommended to use a firewall to mitigate the identified threat.
The architecture of the headlamp system with a firewall is depicted in~Figure~\ref{fig:firewall-security-pattern}. 
A firewall is deployed between the Gateway ECU and the headlamp system.
The goal is to mitigate malicious incoming flows from public elements such as Bluetooth, Cellular or OBD-II connector.

In summary, the machinery can automate the following parts of the ISO 21434 risk assessment:
\begin{itemize}
	\item identification of threats given damage scenarios (threat scenario identification)
	\item identification of the attacker's path to exploit identified threats (attack path analysis)
	\item calculation of the overall level of an attacker's path (attack feasibility rating) given the assigned level to each attack step of an attack path
	\item calculation of the risk value for each identified threat (risk determination)
	\item recommendation of security patterns to mitigate identified threats (risk treatment decision), explicitly suggesting where to deploy them in the architecture 
\end{itemize}

\subsection{Incremental Security Approach}
To illustrate the incremental approach depicted in Figure~\ref{fig:incremental}, assume that there is 
an incremental change in the headlamp system as follows. 

\paragraph{Increment.}
Consider as increment the ability to perform software updates on the Body Control ECU.
It is assumed that at some point during the life-cycle of the headlamp system a software update on the Body Control ECU is performed.
This software update may be performed by, \eg, a wired connection through the OBD-II connector, or through-the-air by using the Bluetooth or Cellular interfaces.

The considered increment may lead to new threats to the headlamp system as new flows are generated between public elements (\eg, OBD-II connector) and the headlamp system.
A potential threat is an attacker injecting malicious code into the Body Control ECU (asset) in an unauthorized fashion. 
This potential threat may lead to loss of integrity of the Body Control ECU due to unauthorized access.

Following the approach illustrated in Figure~\ref{fig:incremental}, the machinery takes the inputs described in Section~\ref{ssec:automating-iso21434} and a potential threat on the Body Control ECU.
The machinery then automatically checks whether the given potential threat is an actual threat by checking possible attacker's paths. 
Three attacker's paths are identified, initiated from the OBD-II connector, Bluetooth and Cellular interface, respectively.

\paragraph{Automated Design Exploration.}
Next, the machinery automatically suggests solutions for mitigating the identified threat.
Two solutions are depicted in Figure~\ref{fig:firewall-secmon-security-patterns}.
It suggested the deployment of a security monitor to mitigate the identified threat.
The goal is to mitigate malicious access from public elements by enforcing access control policies, such as ``only authorized users from public elements may write data into the Body Control ECU''.

Figure~\ref{fig:firewall-secmon-security-patterns-solution1} illustrates a security monitor associated to the Body Control ECU, whereas Figure~\ref{fig:firewall-secmon-security-patterns-solution2} illustrates a security monitor associated to the CAN bus.
Both solutions can mitigate the identified threat.
The main difference is that the former may be deployed by means of software instrumentation, and the latter may be deployed as a physical proxy between the Gateway ECU and the Body Control ECU.

\paragraph{Trade-off Analysis.}
A trade-off analysis should be carried out to help engineers to choose the best suited solution for the system.
For example, the security monitor deployed as a proxy might be more expensive (performance-wise) than the one deployed by means of software instrumentation.
The reason is that the deployed proxy shall intercept each incoming message from the Gateway ECU regardless the destination of the message.
The security monitor associated to the Body Control ECU shall only intercept messages destined to the Body Control ECU itself. 
Hence, based on the performance impact of such solutions, one may choose the security monitor illustrated in Figure~\ref{fig:firewall-secmon-security-patterns-solution1} over the one illustrated in Figure~\ref{fig:firewall-secmon-security-patterns-solution2}.

\begin{figure}
\begin{center}
  \includegraphics[width=0.45\textwidth]{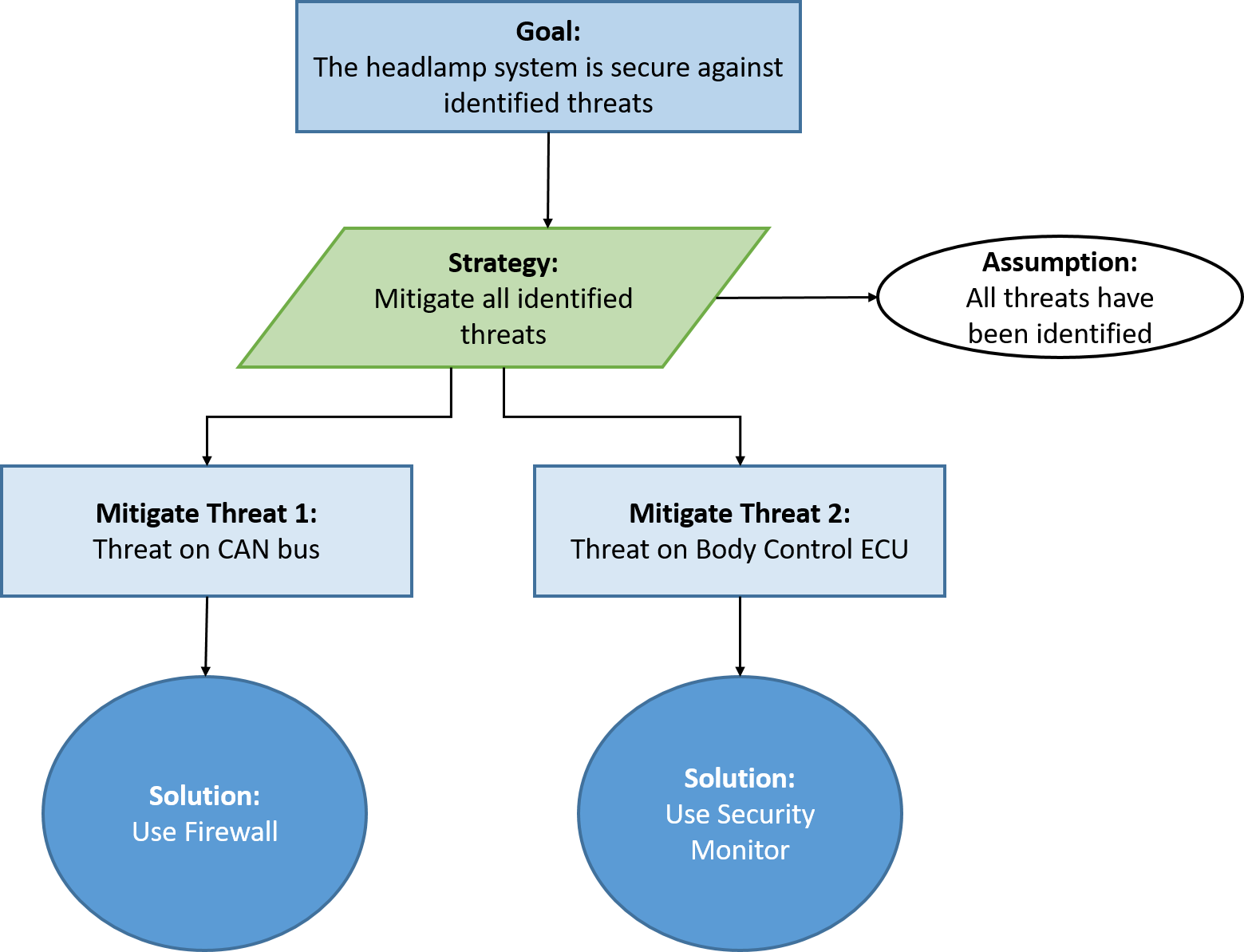}  
\end{center}
  \caption{Security argument in the form of a GSN model derived by the machinery for the headlamp system}
  \label{fig:gsn}
\end{figure}

\paragraph{Automated Security Argument Generation.}
Once a design solution is chosen, the machinery can construct the security argument based on the (previously and newly) identified threats, and the security pattern that have been proposed.
That is, it can construct security arguments in the form of a GSN model.
Figure~\ref{fig:gsn} depicts a (piece of the) GSN model derived from the machinery for the headlamp system.
It includes the identified threat as well as which security pattern can used to mitigate which threat.

\section{Related Work}\label{sec:comparison}

Recent white papers~\cite{whitepaper-hasan}\cite{whitepaper-sembera} also give an overview on ISO 21434.
They both discuss the set of guidelines proposed for securing automotive vehicles.
The white paper~\cite{whitepaper-hasan} focuses on the activities performed in the risk assessment.
It proposes an approach to automotive cybersecurity engineering.
This approach suggests the use of offense and defense mechanisms for helping engineers to implement the guidelines of ISO 21434.
The white paper~\cite{whitepaper-sembera} proposes a layered approach for securing automotive vehicles.
The advantage of this approach is to reduce the probability of an attack's success by providing multi-layered response to attacks for protection, detection, and response.
This paper proposes an incremental approach to enable the continuous certification for automotive vehicles.

ThreatGet is a model-based engineering tool for security analysis~\cite{SadanySK19}\cite{threatget}.
ThreatGet can perform security analysis on system architectures following the ISO 21434 risk assessment.
ThreatGet automatically identifies threats given a path between a source and a target element.
ThreatGet lists possible countermeasure that can be selected by users to mitigate the identified threats. 
The main advantages of the machinery presented here over ThreatGet include that it can automate more parts of the ISO 21434 risk assessment such as both attack feasibility rating and risk determination.
It explicit shows where countermeasures shall be deployed in the system architecture to mitigate identified threats.
It enables the construction of security arguments in the form of a GSN model.
To the best of our knowledge, ThreatGet does not support the generation of such security arguments.
Finally, this paper considers an incremental approach to risk assessment, thus enabling continuous certification.

Goal-oriented approaches, such as GSN~\cite{gsn11standard} and KAOS~\cite{dardenne93scp}, have been used for modeling safety cases. 
Similar to these approaches, several papers~\cite{bagnato12ijsse} have proposed using Attack Trees~\cite{schneier99jst} for modeling risk assessments. 
Extensions include quantitative models for evaluating how defenses can be used to mitigate attacks~\cite{bagnato12ijsse,bistarelli06ares}.
A key difference to the approach proposed here is the focus on automated analysis.
Whereas in the previous work, Attack Defense trees shall be manually constructed, the proposed approach can carry out such analysis in an automated fashion. 


A number of work~\cite{nostro14issrew,safsec.book,meland14ijsse} have proposed using general models encompassing both safety and security concerns. For example, GSN extensions with security features, so that in a single framework, one can express both security and safety~\cite{meland14ijsse}. 
Moreover, our previous work~\cite{kondeva19wosocer,nigam18safsec} proposed mechanisms for extracting security relevant information from 
safety analysis, such as Fault Tree Analysis or Failure Mode and Effects Analysis. 

While this paper is inspired by these previous work, the main focus here is on incremental methods for safety and security co-analysis.
As illustrated by the running example and in our previous work~\cite{dantas20safecomp}, the proposed approach takes into account both safety and security 
during trade-offs to determine the best design increments.

\section{Next Research Directions}\label{sec:next-steps}

The vision of this project is to build an incremental safety and security co-analysis with patterns.
The aim is to develop an incremental process for system safety and security assurance cases using automated methods that incorporate both safety and security reasoning principles.

Some promising steps towards this vision have been made.
Safety reasoning principles have been proposed in~\cite{dantas20iclp}. 
Security reasoning principles with patterns have been partially presented in this white paper and in~\cite{dantas20safecomp}.

Some more research is needed to fully realize the vision. 
The next steps toward this vision include:
\begin{enumerate}
	\item \textbf{Make Assumptions Explicit:}  
	A safety case often contains implicit assumptions (\eg, decisions based on own experience), \eg, assumptions on channel independence or lack of side-channel attacks.
  Often no evidence is collected validating such implicit assumptions which may lead to implementation errors.
	The purpose is to make such assumptions more explicit, \eg, what are the assumptions needed for choosing a particular safety or security pattern?
	The next step is to augment the reasoning principles rules~\cite{dantas20iclp}\cite{dantas20safecomp} with explicit assumptions to enable a more precise analysis w.r.t., \eg, pattern selection. 

	\item \textbf{Improve the trade-off analysis:}
	Safety and security co-analysis may lead to interrelations.
	There can be conflicts, synergies or no conflicts between safety and security. 
    For example, implementing a firewall needed for security reasons might lead to new faults (safety) as the firewall might incorrectly blocks messages from the system.
    Upon finding out the trade-offs, safety and security engineers will possibly decide what actions to take to solve conflicts or synergies.
    Engineers shall take into account aspects, \eg, performance or safety, to reach a consensus w.r.t. solutions for the system.
    The next step here is to investigate directions on how to integrate such aspects into the trade-off analysis in a (semi-) automated fashion.
    As a result, one can provide a more holistic trade-off analysis with multiple aspects that can assist engineers to decide the best design solution for the system.

    \item \textbf{Tooling.} Currently, the machinery is based on the logic programming engine DLV~\cite{dlv} for automating safety and security reasoning principles such as pattern recommendation.
    As mentioned earlier, performance is one of the aspects to be integrated into the trade-off analysis.
    To this end, reasoning principles with timing may be specified so that one can analyze possible trade-offs regarding the runtime overhead caused by solutions.
    DLV might not be suitable for specifying such reasoning principles with timing.
    Moreover, reasoning principles on the probability of faults occurrence may be specified to enable a more precise analysis w.r.t. pattern selection. 
    The next step is to investigate what tools can be used to specify such reasoning principles.
    For example, one could combine DLV with SMT-solvers to enable the specification of reasoning principles with fault probabilities.

   \item \textbf{Integration into Model-Based Engineering Approaches.}
    One can imagine the machinery (including DSL, patterns, and reasoning principles) as a back-end for model-based engineering approaches.
	That is, it can be used as a process inside a model-based engineering tool to provide rigorous safety and security assessment to the given system architectures.
	Therefore, the next step here is to integrate the machinery into the model-based engineering tool AutoFOCUS3~\cite{af3}.
	This integration leads to a number of benefits include (a) assisting engineers to operationalize the process of providing rigorous assessment to system architecture, and (b) improving the usability of the machinery by providing a user-friendly graphical interface tool for engineers.
\end{enumerate}	

\section{Conclusions}\label{sec:conclusion}

This paper describes the ISO 21434 standard for automotive security focusing on the activities and artefacts that shall be produced by engineers. 
The paper goes beyond the ISO 21434 by describing an incremental security engineering approach. 
It illustrates by example how this approach can support engineers in constructing arguments for item security claims. 
Finally, it points to multiple research directions to realize the proposed incremental approach in an operational setting.

The ISO 21434 is a step forward towards increasing vehicle security as it addresses some of the key challenges upcoming with the increase adoption of ICT in vehicles.
However, without adequate automated methods, it may fall short given the tight deadlines in placing vehicles in the market. 
This pressure may cause engineers to cut corners by overlooking (implicit) assumptions or not taking into account the trade-offs between safety and security thus leading to insecure vehicles.

Therefore, the next years will require a co-joint effort between research institutes, industry and certification bodies to develop the framework needed, \ie, methods and processes, to efficiently derive appropriate security arguments supported by comprehensive evidence. 
\phantomsection
\bibliographystyle{plain}
\balance
\bibliography{bib}


\end{document}